\documentstyle[12pt,epsf,epsfig]{article}
\def\beq{\begin{equation}}
\def\eeq{\end{equation}}
\def\bea{\begin{eqnarray}}
\def\eea{\end{eqnarray}}
\def\ba{\begin{array}}
\def\ea{\end{array}}

\def\calo{{\cal O}}

\newcommand{\NPB}[3]{\emph{ Nucl.~Phys.} \textbf{B#1} (19#2) #3}   
\newcommand{\PLB}[3]{\emph{ Phys.~Lett.} \textbf{B#1} (19#2) #3}   
\newcommand{\PRD}[3]{\emph{ Phys.~Rev.} \textbf{D#1} (#2) #3}

\newcommand{\PR}[3]{\emph{ Phys.~Rep.} \textbf{#1} (19#2) #3}

\def\gappeq{\mathrel{\rlap {\raise.5ex\hbox{$>$}}
{\lower.5ex\hbox{$\sim$}}}}
\def\permil{$\%\raise.20ex\hbox{$_0$}}
\def\lappeq{\mathrel{\rlap{\raise.5ex\hbox{$<$}}
{\lower.5ex\hbox{$\sim$}}}}
\begin{document}
\topmargin -1.0cm
\oddsidemargin -0.8cm
\evensidemargin -0.8cm
\pagestyle{empty}
\begin{flushright}
\end{flushright}
\vspace*{5mm}
\begin{center}
{\Large\bf Ultraviolet dependence of 
Kaluza-Klein effects  
}\\
\vspace{0.5cm}
{\Large\bf on electroweak observables}\\
\vspace{2cm}
{\large\bf M. Masip}\\
\vspace{.8cm}
{{\it Department of Physics and Astronomy}\\
{\it University of Iowa}\\
{\it Iowa City, Iowa 52242, USA}\\
\vspace{.8cm}
{\it Departamento de F\'\i sica Te\'orica y del Cosmos}\\
{\it Universidad de Granada}\\
{\it E-18071 Granada, Spain}\\}

\end{center}
\vspace{1.4cm}
\begin{abstract}

In extensions of the standard model (SM) with $d$ extra dimensions 
at the TeV scale the virtual exchange of Kaluza-Klein (KK) 
excitations of the gauge bosons gives contributions that
change the SM relations between electroweak observables.
These corrections are finite only for $d=1$; for $d\ge 2$ 
the infinite tower of KK modes gives a divergent contribution 
that has to be regularized introducing a cutoff (the string
scale).
However, the ultraviolet dependence of the KK effects
is completely different if the running of the couplings with 
the scale is taken into account.
We find that for larger $d$ the number of excitations at each 
KK level increases, but their larger number is compensated by 
the smaller value of the gauge coupling at that scale. As a 
result, for any number of extra dimensions the exchange of 
the complete KK tower always 
gives a finite contribution. We show that {\it (i)} for 
$d=1$ the running of the gauge coupling decreases a $14\%$ 
the effect of the KK modes on electroweak observables; 
{\it (ii)} in all cases more than $90\%$ of the total effect
comes from the excitations in the 
seven lowest KK levels and is then independent
of ultraviolet physics.

\end{abstract}

\vfill

\eject
\pagestyle{empty}
\setcounter{page}{1}
\setcounter{footnote}{0}
\pagestyle{plain}


\section{Introduction}

Theories with large extra compact dimensions 
offer new ways to accommodate the hierarchies observed 
in particle physics \cite{ark98}. They can be realized
within a string theory \cite{ant90}, the only candidate  
for a consistent description of quantum gravity.
Phenomenologically, 
the nonstandard effects that extra dimensions imply
can be supressed to acceptable levels by their
size and their properties (for example, an
extra dimension could be invisible to some fields and
not to others).
In particular, the possibility of compact dimensions 
at scales $M_c \equiv R^{-1} = \calo(1\; {\rm TeV})$
has recently received a lot of attention. 
Such dimensions, with $M_c$ below the string scale, 
could explain gauge unification \cite{die98} or supersymmetry 
(SUSY) breaking \cite{pom98}.

The momentum along a compact dimension is quantized
in units of $M_c$, and a higher dimensional field  
is seen in four dimensions as an 
infinite tower of Kaluza-Klein (KK) modes of mass
$nM_c$. Let us consider a theory where the 
gauge bosons can propagate along the extra dimensions.
Its low-energy limit will resemble an extension of 
the standard model (SM) with massive replications of 
all the gauge fields. The virtual exchange of these fields 
will introduce corrections of order
$M_W^2/M_c^2$ to the interactions mediated by the
$Z$ and $W$ bosons and will change the SM relations 
between electroweak observables \cite{yam99}.
For $d=1$ and compactification on a circle
there are two states at each KK level $n$. The exchange of
the complete tower of resonances gives then corrections
proportional to 
\begin{equation}
\sum_{n=1}^{\infty} {2M_W^2\over n^2 M_c^2} = 
{\pi^2 M_W^2\over 3 M_c^2}\;.
\end{equation}
The main contribution in this sum is due to the 
lightest resonances and thus the result is not
dependent on ultraviolet details. Actually, the effect 
of the whole KK tower mimics the effect of a single 
resonance with $M^2=3M_c^2/\pi^2$.

However, the fact that there is an infinite tower
of massive modes instead of just one state is specially
relevant for more than one extra dimension. 
For $d\ge 2$ the number of excitations at each KK level
grows with $n$ (see below) and the contribution from the 
heavy modes diverges.
This divergence, related to the non renormalizability 
of higher dimensional gauge theories, implies that 
at larger scales the theory 
has to be embedded in a more 
fundamental framework. In particular, in string theory 
one would expect an exponential supression of the gauge
coupling of the
KK modes with masses above the string scale \cite{ant94}, 
which would act as an effective cutoff.

From the previous argument it seems to follow that, 
even if there is a gap between the compactification
and the string scales (as suggested by gauge unification),
for $d\ge 2$ the virtual exchange of KK modes is 
always going to 
be dominated by contributions near the cutoff, where 
genuine KK effects would be mixed with string effects.
We will show that this is not the case. 
The heavy excitations of the gauge bosons have weaker 
couplings to fermions than the lighter modes, and for any
value of $d$ the main contribution comes
from the exchange of the fields in the lowest KK levels.

\section{Running of the gauge couplings}

The evolution of the gauge couplings with the scale depends
essentially on the total number of excitations at each KK level.
Let us consider a model with 
$d$ compact dimensions with common radius $R=M_c^{-1}$.
The momentum along the extra dimensions is given by the 
vector $\vec n M_c=(n_1,n_2,...,n_d) M_c$, where $n_i$ can 
take any integer value. This momentum is seen in four 
dimensions as the mass 
\begin{equation}
M_n=\sqrt{n_1^2+...+n_d^2} M_c\equiv nM_c\;.
\end{equation}
Each KK mode {\it occupies} a point in a $d$--dimensional
reticle. To calculate the number $N(\mu)$ of KK states with a mass
smaller than $\mu=nM_c$ we can approximate this discrete 
distribution by the averaged (continuous) one, and then 
$N(\mu)$ is just the volume of a $d$--dimensional
sphere of radius $n$:
\begin{equation}
N(\mu)={\pi^{d/2}\over \Gamma({d\over 2}+1)} \Big( {\mu\over M_c}
\Big)^d\;,
\end{equation}
where the Gamma function above satisfies
$\Gamma({1\over 2})=\pi^{1/2}$, 
$\Gamma({d\over 2}+1)={d\over 2}\Gamma({d\over 2})$ and
$\Gamma(d+1)=d!$, being $d$ a positive integer.

The KK approach provides a simple framework to understand
the running of the gauge couplings in 4+$d$ dimensions. 
At the lowest order the running is given by 
the one-loop contributions to the vector boson
selfenergy. At each scale $\mu$, we have to include in 
the loop the KK modes lighter than $\mu$\footnote{
The authors in \cite{die98} show that for $d=1$ the 
one-loop corrections to the gauge couplings in the 5D theory
almost coincide with the running in the truncated theory 
where the heavier modes have been cut off. The regularization
dependence and KK threshold effects have
been discussed in \cite{kub99}.}. Due to this 
multiplicity the renormalization group equations become
\begin{equation}
\mu {{\rm d} \alpha^{-1}\over {\rm d}\mu}  = -{Q\over 2\pi} 
\longrightarrow
\mu {{\rm d} \alpha^{-1}\over {\rm d}\mu}  = -{Q\over 2\pi} 
{N(\mu)} \;, 
\end{equation}
where $\alpha=g^2/(4\pi)$, 
$Q={1\over 6} T(S) + {2\over 3} T(F) - {11\over 3} C(V)$,
$S$, $F$ and $V$ stand respectively for scalar, fermion 
and vector fields, $T(\Phi)\delta_{AB}= {\rm Tr}(T^AT^B)$ for
$\Phi=S,F$, and $C(V)\delta_{AB} = f_{ACD}f_{BCD}$.
The first equation (the case with no KK resonances) 
gives a logarithmic dependence of $\alpha^{-1}$ 
on $\mu$, whereas the second one predicts a much faster 
power-law behaviour:
\begin{equation}
\alpha^{-1}(\mu)=\alpha^{-1}(M_c)-{Q\pi^{d/2-1}\over
2d\; \Gamma({d\over 2}+1)} \Big( {\mu^d\over M_c^d} -1 
\Big)\;.
\end{equation}

Let us focus on the $SU(2)_L$ gauge coupling
of a non-SUSY extension of the SM with $d$ extra dimensions 
where only gauge and Higgs (but not quark and lepton) fields 
propagate. Taking into account the additional degrees of freedom
of a vector field in $4+d$ dimensions, we find 
$Q_L=(1+2d-44)/6$. This negative beta function will make
$\alpha_L$ a decreasing funtion of the scale. From Eq.~(5) it
follows that, for large enough $\mu$, $\alpha$ decreases like
$1/\mu^d$. In Fig.~(1) we plot the running of 
$\alpha_L$ for $M_c=1$ TeV and different values of $d$.

Two comments are in order. First, 
the $U(1)_Y$ gauge coupling has a positive beta function 
(due to Higgs contributions only) and
it will grow with the scale. However, $\alpha_Y$ has initially
a smaller value than $\alpha_L$, and both couplings 
will then coincide at a larger scale (around $20M_c$ for
$d=1$ \cite{die98}). This suggests unification into a larger
gauge group, like $SU(3)_L\times SU(3)_R$ or $SU(5)$, 
where Yang-Mills interactions would provide for a negative 
beta function and a weaker coupling at higher scales.
Second, we would like to mention the SUSY case. Here
one needs to take into account the different 
field content of a fermion in higher dimensions: 
a minimum of four components if
$d=1,2$ or eight components for larger $d$ \cite{soh85}. 
In the first
two cases the matter content can be accommodated in
complete 4--dimensional
multiplets of $N=2$ SUSY: vector superfields
$V$ will come together with a chiral superfield $\Sigma$ 
in the adjoint representation of the gauge group, whereas 
the Higgs doublets will come in pairs $H,H'$ of 
chiral superfields. The generic SUSY 
beta function with $Q=T(chiral)-3C(vector)$ becomes now
$Q=T(H)+T(H')+T(\Sigma)-3C(V)=2T(H)-2C(V)$, with
$Q=-3$ for $SU(2)_L$. We include the $d=1,2$ cases in
Fig.~(1). For larger $d$, the higher dimensional fields
complete hypermultiplets of $N=4$ SUSY,  
including a vector plus three chiral superfields all in
the adjoint representation of the group. In such 
scenario one is 
forced to extend the $SU(2)_L$ gauge symmetry 
to have Higgs doublets. Moreover, $N=4$ theories 
are finite and the beta function vanishes \cite{soh85}.
However, the spectrum of zero modes as well as the
interactions of 
the fermion fields living in 4 dimensions will break the 
$N=4$ SUSY and at the two-loop level will give nonzero 
beta functions with a model dependent sign \cite{mas00}.

\section{Effect of the KK excitations on electroweak observables}

As shown in the previous section, the gauge couplings
decrease like a power law with the scale, implying
that heavier KK modes couple to fermions with smaller
strength. This will give an important correction to the
cumulative effects of the KK tower on electroweak observables.

In particular, let us consider the Fermi coupling measured
in muon decays. Its definition will include now the exchange of  
the $W$ boson and its excitations:
\begin{equation}
{\sqrt{2}\over \pi} G_F = {\alpha_L\over M_W^2}
+ \sum_{n} {\alpha_L\over n^2 M_c^2}\;.
\end{equation}
The sum above can be approximated 
by an integral. The number 
of modes ${\rm d} N(\mu)$ at the $n$ level 
({\it i.e.}, with a mass in the interval $\mu + {\rm d}\mu
=(n + {\rm d} n )M_c$; notice that $n=\sqrt{n_1^2+...+n_d^2}$ 
is not necesarily an integer number) is
\begin{equation}
{\rm d} N(\mu)={2\pi^{d/2}\over \Gamma({d\over 2})} 
{\mu^{d-1}\over M_c^d} {\rm d} \mu\;.
\end{equation}
Then
\begin{equation}
\sum_{n} {\alpha_L\over n^2 M_c^2}\longrightarrow
I(\Lambda) \equiv \int_{M_c}^{\Lambda} 
{\alpha_L(\mu)\over \mu^2} {\rm d} N(\mu)\;,
\end{equation}
where we have introduced a cutoff mass $\Lambda$. 
Since the number ${\rm d} N(\mu)$ of excitations with 
mass $\mu$ grows proportional to $\mu^{d-1}$, if the 
running of $\alpha_L$ is neglected the integral $I(\Lambda)$
converges only for $d=1$ and diverges like $\log \Lambda$ 
for $d=2$ and like $\Lambda^{d-2}$ for larger $d$.
However, $\alpha_L$ decreases proportional to $1/\mu^d$ 
at large $\mu$. Taking into account the weaker coupling
to quarks and lepton of the heavy modes, the
dependence of the integral on the cutoff vanishes like
$1/\Lambda^2$ for all $d$ and the result is
always finite.

In Fig.~(2) we plot the value of 
$I(\Lambda)$ for $M_c=1$ TeV and different values of
$d$. We also plot $I_0(\Lambda)$, the result that would 
be obtained neglecting the running of the gauge 
coupling ({\it i.e.}, replacing $\alpha_L(\mu)$ by 
$\alpha_L(M_c)$ in Eq.~(8)). 
The figure shows
that the running is very effective in suppressing the
effect of heavier modes. In particular, we find that more than 
$90\%$ of the total correction to $G_F$ comes from the 
excitations in the 6 lowest KK levels for $d=1$ and $d=6$ or
from the 7 lowest levels for $d=2$. The KK effects
are then insensitive to physics beyond $\approx 7M_c$.

In Fig.~(3) we plot the ratio 
$I(\Lambda)/I_0(\Lambda)$, which gives the correction 
to KK effects due to the running of $\alpha_L$.
In the case with $d=1$, the most extensively
considered in the literature \cite{yam99}, the running 
reduces the KK effects on $G_F$ (and the bound on $M_c^2$) 
in just a $14\%$ (or an $11\%$ for a cutoff $\Lambda=20 M_c$). 
However, for $d=2$ the correction is a factor of 0.63
if $\Lambda = 10 M_c$ and a factor of 0.09 for $d=6$ with
$\Lambda = 4M_c$.

\section{Conclusions} 

Bounds on new physics from precision electroweak data are
usually obtained from analysis that combine 
new physics at the tree level with standard model 
effects at one loop. We have shown that in models 
with extra dimensions at the TeV scale this procedure is
in general not justified. 
At the one-loop level the gauge couplings experience 
${\cal O}(1)$ power-law corrections that decouple the heavy 
KK modes. We find that for $d=1$ the 
running of $\alpha_L$ accounts for 
a small reduction (around $14\%$) of the tree-level KK 
effect on $G_F$, but for $d\ge 2$ it
makes finite an effect that diverges at the tree level.

The divergent tree-level result is just expressing 
that these (non-renormalizable) higher dimensional 
extensions of the SM should 
be embedded at larger scales in a string theory.
Our result, however, shows that including the one-loop 
running of the gauge couplings it is not necessary
to introduce an ultraviolet cutoff in order to cure
these pathologies. KK effects on electroweak observables
are insensitive to the value of the ultraviolet cutoff or
to the way the theory is embedded on the fundamental 
theory if the larger scale is $\ge 7M_c$.

\section*{Acknowledgements} 

The author thanks Yannick Meurice and Hallsie Reno for 
discussions. 
This work was supported by CICYT under contract AEN96-1672, 
by the Junta de Andaluc\'\i a under contract FQM-101 and
by a grant from the MEC of Spain.

\newpage

\setlength{\unitlength}{1cm}
\begin{figure}[htb]
\begin{picture}(8,19.5)
\epsfxsize=21.cm
\put(-1.1,-15.0){\epsfbox{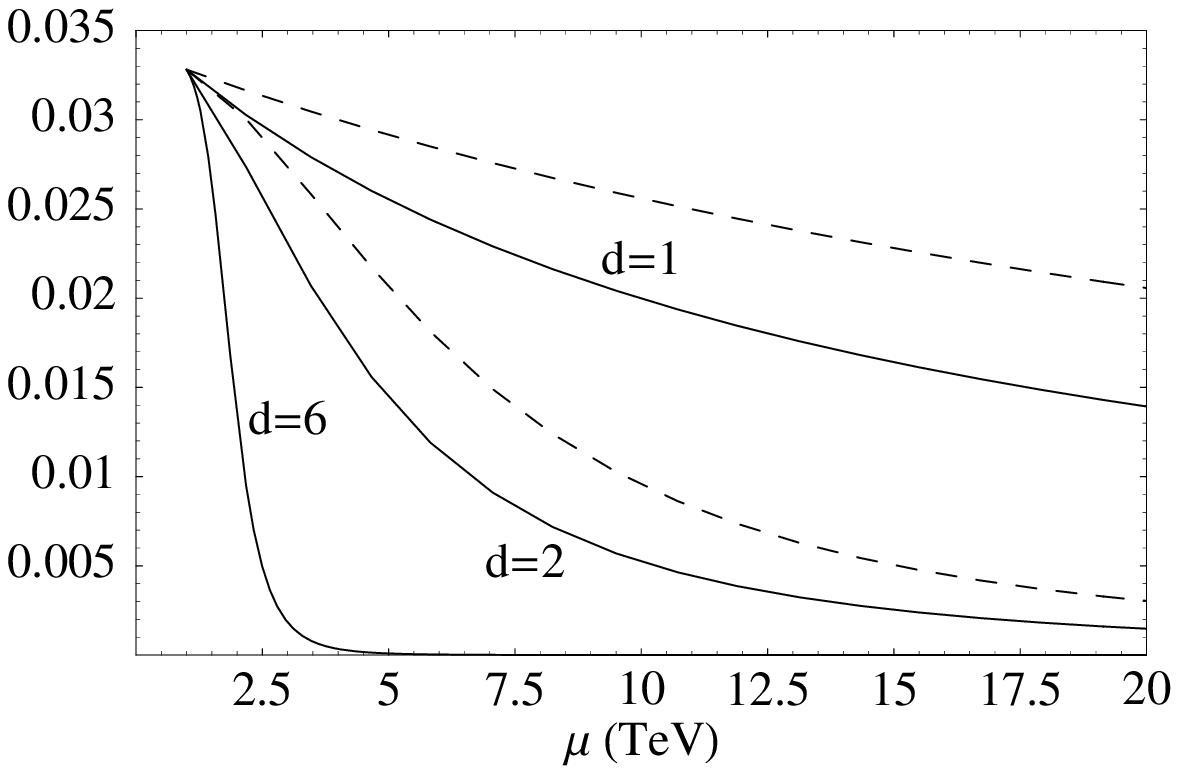}}
\end{picture}
\caption{
Running of $\alpha_L$ for $M_c=1$ TeV and different values
of $d$. Dashed lines for $d=1,2$ correspond to the minimal 
SUSY extensions.
\label{Fig. l}}
\end{figure}

\newpage

\setlength{\unitlength}{1cm}
\begin{figure}[htb]
\begin{picture}(8,19.5)
\epsfxsize=21.cm
\put(-1.37,-6.5){\epsfbox{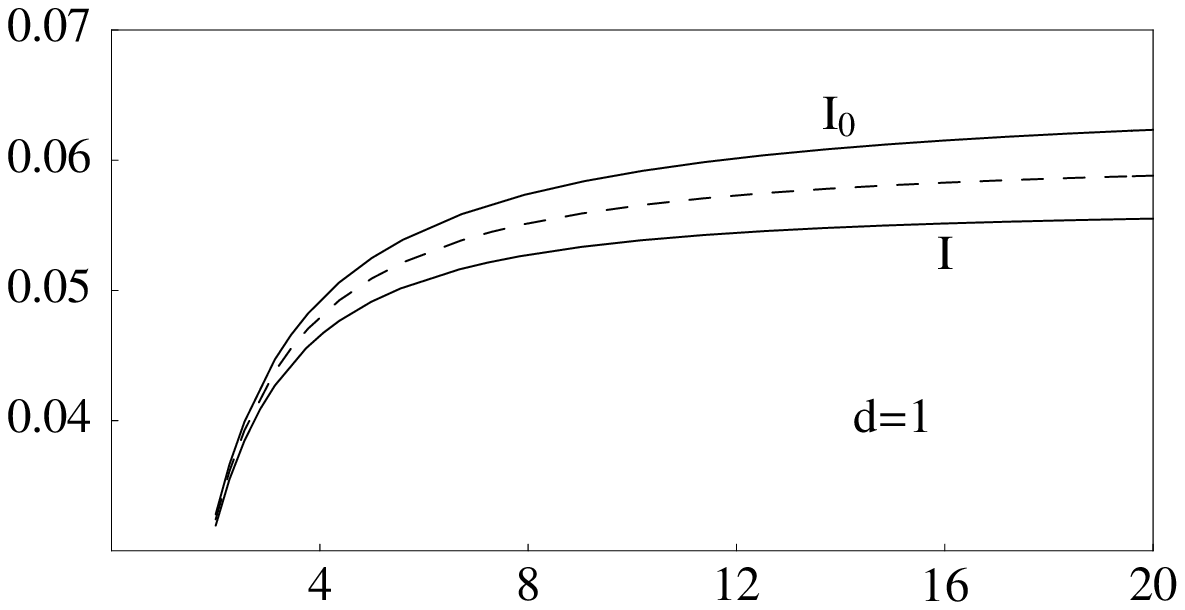}}
\put(-1.1,-13.0){\epsfbox{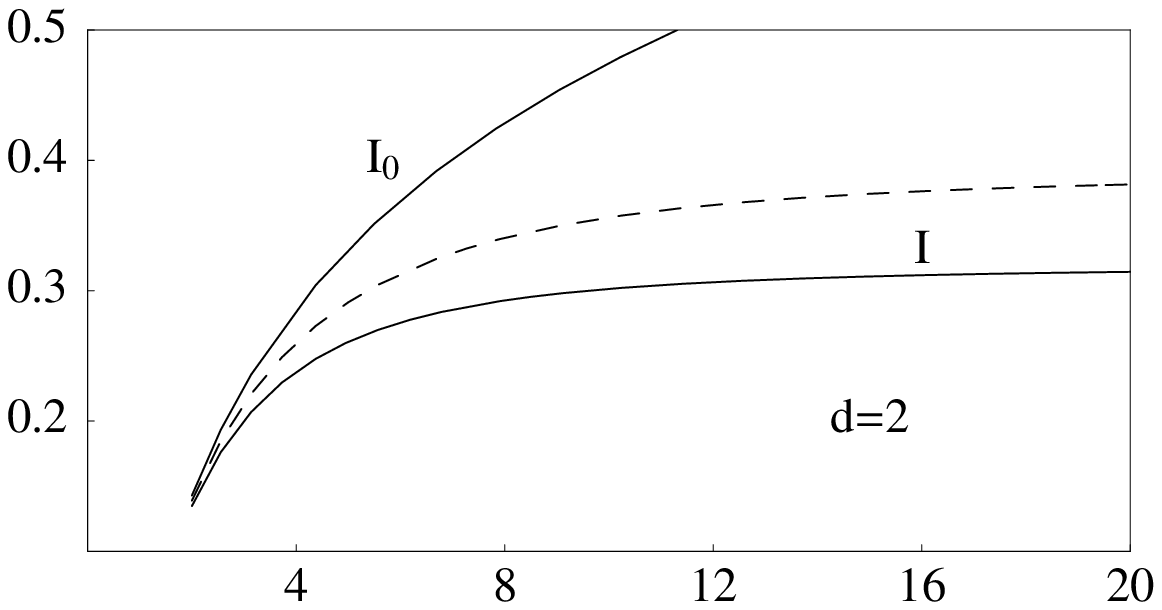}}
\put(-1.,-19.5){\epsfbox{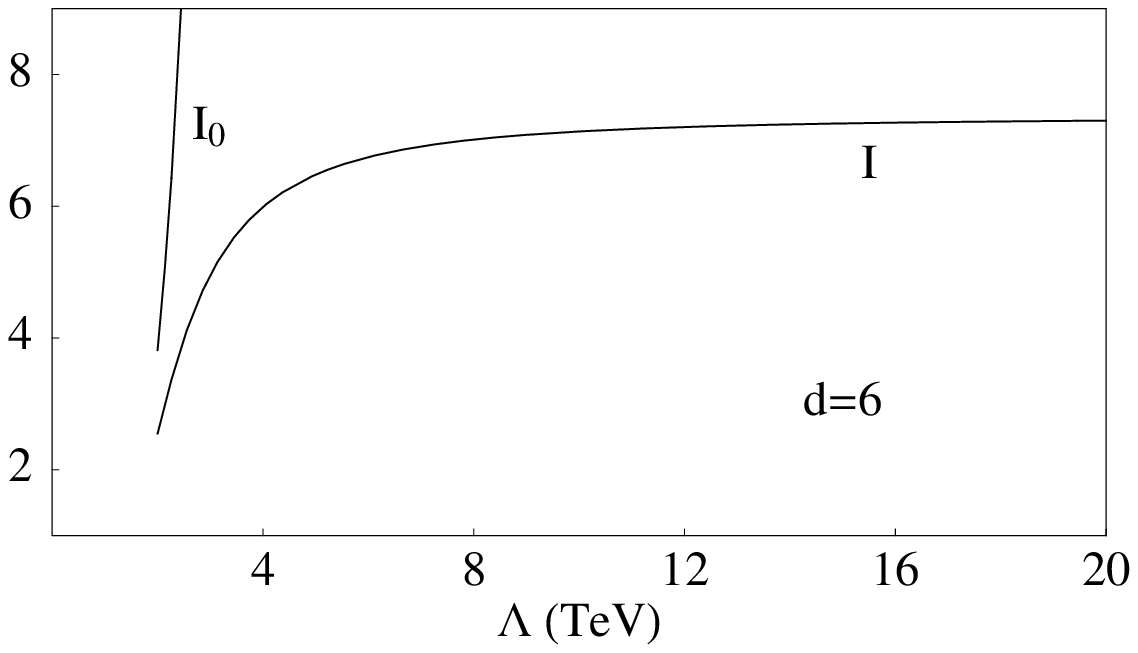}}
\end{picture}
\caption{
$I(\Lambda)$ and $I_0(\Lambda)$ ({\it i.e.,} the 
value of $I(\Lambda)$ neglecting the running of $\alpha_L$)
for $M_c=1$ TeV and different values of $d$. Dashed lines 
for $d=1,2$ correspond to the minimal SUSY extensions.
\label{Fig. 2}}
\end{figure}

\newpage

\setlength{\unitlength}{1cm}
\begin{figure}[htb]
\begin{picture}(8,19.5)
\epsfxsize=21.cm
\put(-1.1,-15.0){\epsfbox{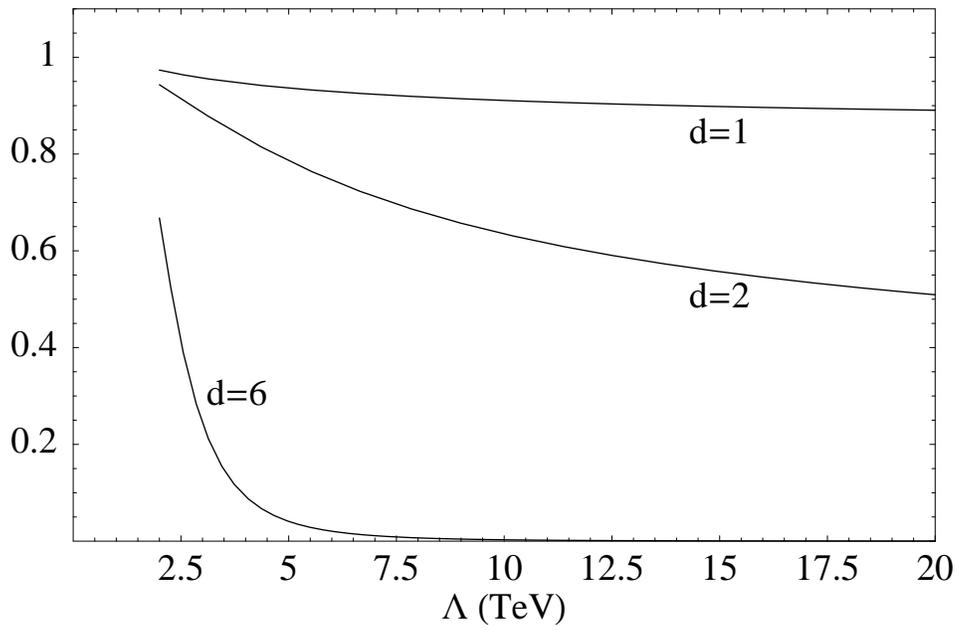}}
\end{picture}
\caption{
Ratio $I(\Lambda)/I_0(\Lambda)$ for $M_c=1$ TeV and different 
values of $d$. 
\label{Fig. 3}}
\end{figure}

\end{document}